# Who Governs Data in the AI Era? A Computational Analysis of the U.S. Privacy Workforce in Job Postings

Ramazan Yener[1] Muhammad Hassan[2] and Masooda Bashir[3]

[1]University of Illinois Urbana- Champaign, USA

[2]University of Illinois Urbana- Champaign, USA

[3]University of Illinois Urbana- Champaign, USA

## Abstract

Privacy protection now spans legal, technical, and managerial duties, and demand for privacy professionals is growing across sectors. However, little is known about how employers define these roles. We analyze 1,143 U.S. privacy job postings from LinkedIn and Indeed. We examine job titles, salaries, competencies, certifications, education, experience, regulatory references, and AI-related language by using rule-based text mining. We also apply Topic Modeling (BERTopic) to the same postings and identify 18 latent themes which we grouped them into four categories. Our findings show that privacy roles are hybrid and they combine legal knowledge, technical skills, and interpersonal competence. Artificial intelligence appears in more than half of postings, with AI language spread across compliance, legal, governance and security themes. Our research indicates that AI governance responsibilities are often embedded within existing privacy roles, contributing to the rise of hybrid positions alongside dedicated AI governance roles.

## 1. Introduction

Organizations now collect large amounts of sensitive data which includ health records, behavioral traces, biometric identifiers, and AI-generated inferences. This creates business value but also raises risks such as privacy violations, lawsuits, and reputational harm. Public concern about data misuse has grown alongside these developments (Cisco Systems 2025). The rise of AI introduces further risks, as AI systems can infer sensitive traits and make automated decisions affecting privacy and fairness. An industry survey found that 68 percent of companies report increased demand for professionals skilled in AI governance and data ethics (IAPP 2024a).

We examine who organizations hire to do that work. Previous research on the security workforce has started to use job advertisements as evidence. For example, Ozyurt and Ayaz (2024) analyzed 9,407 cybersecurity job postings from Indeed using descriptive analysis and topic modeling, identifying ten job titles and twenty-three skill sets expected in that field. Earlier work looked at the security occupation more directly and Ludbey, Brooks, and Coole (2020) surveyed and interviewed corporate security staff across four Australian organizations and found a progression ceiling that placed corporate security within the technostructure as a specialist function with limited executive reach. We bring together the demand-side approach of the first study and the occupational focus of the second, applying them to the privacy workforce, which has not yet received comparable analysis.

Governance has increasingly become a distinct focus within security research. For example, Radanliev (2024) explores how the interaction of technology and policy shapes digital security by design across various sectors. Szadeczky and Bederna (2025) also assess AI systems across different risk tiers alongside the regulatory frameworks required to manage them. Both of these studies integrate regulatory and ethical oversight directly into the broader security landscape. Our research empirically identifies which specific labor market roles assume this oversight and details the expectations employers place on them.

We use the term AI governance to refer to the organizational and policy structures that oversee AI systems and manage the risks they create. According to the IAPP, AI governance consists of the laws, policies, frameworks, and organizational practices that guide AI toward responsible and compliant use (IAPP 2024b). As AI systems increasingly process personal data and generate privacy risks through inference and profiling, privacy becomes directly relevant (Schwartz, Jones, and Chaudhry 2024).

We examine the privacy workforce as it is defined by employers. In job postings, privacy is often used as a broad term that includes data protection, regulatory compliance, security and risk management, data governance, AI governance, and data ethics. Our dataset reflects this industry usage. Hence, claims we make about the definition of privacy work refer to how employers describe privacy roles not about to normative meanings of privacy.

Regulations like GDPR, HIPAA, CCPA, and the EU AI Act establish specific requirements for handling personal data. Following these rules is challenging for organizations, and failures reduce public trust in digital systems (Samarin 2024). Data protection has therefore become central to reputation and business performance (Gstrein and Beaulieu 2022; Rommetveit and van Dijk 2022). In the United States, more than 20 states now have comprehensive privacy laws, each granting consumers new rights (Botero and IAPP Westin Research Center 2025). Therefore, companies need privacy professionals to handle all these legal compliance, AI governance, and technical implementations.

Despite this growth, academic research on the privacy profession from a workforce perspective remains limited. Existing studies mainly focus on specific roles, especially privacy engineers by using interviews and surveys (Kilhoffer, Wilder, and Bashir 2024; Samarin et al. 2025; Yener, Hassan, and Bashir 2025). Additionally, industry reports provide broader frameworks and salary benchmarks (IAPP 2024a; IAPP 2024c), but they do not examine the day-to-day expectations described in job postings. As a result, relatively little is known about how private-sector employers define privacy roles and the types of privacy professionals they seek.

Privacy outcomes depend on how people, processes, and systems interact. They are also shaped by regulatory requirements and commercial pressures. Technical controls and policy statements alone do not determine these outcomes. Iwaya et al. (2023) show that privacy engineering is embedded in organizational practice and cannot be reduced to a technical specialty. Samarin et al. (2025) describe privacy professionals as intermediaries who translate between legal and technical domains. Employers look for a mix of technical, interpersonal, and organizational competencies. Prior work also shows that implementing privacy obligations requires translating broad regulatory principles into concrete engineering practices such as design patterns and technical building blocks (ENISA 2022; EDPS 2014).

To examine these expectations, we conducted a computational analysis of 1,143 U.S. privacy job postings collected from LinkedIn and Indeed. We used two complementary analytical approaches on the same corpus. First, we applied rule-based text mining and descriptive statistics to identify explicitly stated requirements, including skills, qualifications, regulations, and other job characteristics. We then used BERTopic to examine the latent thematic structure of the same job decsriptions and identify broader areas of privacy work. These approaches show both what employers request and how those expectations cluster across the privacy workforce. Our study findings suggest that organizations often integrate AI governance into existing privacy roles, with positions frequently combining responsibilities for AI accountability, risk management, compliance, and coordination across legal, technical, and governance functions.

We address seven research questions using two complementary analytical approaches. The rule-based text mining and descriptive analysis examine the most common job titles (RQ1), the skills and competencies employers seek (RQ2), educational qualifications and experience requirements (RQ3), certifications (RQ4), which regulations, frameworks, and standards are mentioned (RQ5), and what compensation patterns are observable (RQ6). BERTopic is used to identify latent themes in the job descriptions and examine how these themes characterize the privacy profession (RQ7). Across these analyses, we examine how employers connect privacy and AI governance in hiring expectations and what competency profiles are associated with these roles. These research questions provide an empirical view of how the privacy workforce is defined in the U.S. job market. The findings may also inform educators and curriculum designers as they prepare students for privacy roles that increasingly incorporate AI governance responsibilities.

## 2. Background and Related Work

### 2.1 The Rise of the Privacy Profession

Samarin et al. (2025) describe privacy engineers as cross-functional collaborators who coordinate privacy reviews and help align system design with regulatory requirements.

Privacy engineering itself has been defined as the systematic process of identifying and addressing privacy concerns during the development of sociotechnical systems (Gurses and Del Alamo 2016). Earlier work by Gürses, Troncoso, and Diaz (2011) also highlights the importance of engineering expertise, showing how technical knowledge can shape the translation of privacy concerns into system requirements and design decisions.

As privacy became both a compliance requirement and a strategic concern for organizations, they have been creating dedicated privacy roles. Bamberger and Mulligan (2011) showed that corporate privacy officers became important organizational actors, helping translate formal legal requirements into internal practices. The GDPR further accelerated this development. Heimes and Pfeifle (2016) projected roughly 75,000 Data Protection Officer positions worldwide, and within a year of the regulation taking effect, the number of privacy professionals in Europe had reached about 500,000 (Fennessy 2019; Rodriguez 2018). The profession also expanded more broadly, with IAPP membership increasing from 30,000 in 2017 to more than 50,000 globally by 2019 (IAPP and EY 2017; IAPP 2019). Organizations also report benefits from these investments: 96% of respondents in the Cisco Data Privacy Benchmark Study 2025 said that the benefits of privacy investments outweighed the costs (Cisco Systems 2025), while TrustArc (2023) reported continued growth in dedicated privacy teams across organizations of different sizes.

AI has since become the leading challenge in privacy management. AI is the leading challenge named by privacy teams, with 47 percent identifying it as their biggest hurdle (TrustArc 2025), and by 2025, 77 percent reported active work on AI governance (Sentinella et al. 2025). More than half of privacy functions (55 percent) are directly responsible for AI governance, and 69 percent of chief privacy officers report AI-related oversight duties (IAPP 2024a). ISACA describes a new triad in which privacy, cybersecurity, and legal functions jointly manage AI risk, with privacy operating as a distinct pillar alongside security (Demann 2025).

### 2.2 Competencies and Roles of Privacy Professionals

Research on the competiencies and responsibilities of privacy professionals remains limited. Most existing studies rely on interviews or surveys and focus primarily on privacy engineers (Kilhoffer, Wilder, and Bashir 2024; Samarin et al. 2025; Yener, Hassan, and Bashir 2025; Iwaya, Babar, and Rashid 2023). In the literature we reviewed, we found no peer-reviewed research that examines privacy-related job postings at scale. Addressing this gap is a central goal of the present study. These studies describe a multidisciplinary profession. Kilhoffer et al. (2024) show that privacy engineers act as liaisons between legal and engineering teams and as educators, and Yener et al. (2025) report that privacy professionals hold several roles at once, including trainer, manager, bridge, implementer, and advocate. Additionally, Samarin et al. (2025) identify cross-functional stakeholder management, risk and threat management expertise, and the centrality of privacy reviews as core skills.  These findings suggest that privacy professionals are taking on a broader range of responsibilities. More than 80% or privacy professionals report responsibilities beyond their primary privacy role, while 68% have assumed additional responsibilities related to AI governance (IAPP 2024a; Walczak et al. 2025). It remains unclear if this expansion represents a lasting change in the privacy profession or a temporary overlap between emerging and existing responsibilities.

### 2.3 Job Postings as Workforce Data

Job postings provide a direct and relatively current view of the skills employers seek. Earlier studies often relied on manual coding of job advertisements, which limited the size of the datasets that could be analyzed (Todd, McKeen, and Gallupe 1995; Khan and Du 2018). More recent research has used automated text mining and topic modeling to study job postings in areas such as Industry 4.0, business intelligence, big data analytics, and cybersecurity (Pejic-Bach et al. 2020; Debortoli, Muller, and vom Brocke 2014; Wowczko 2015; Persaud 2021; Grybauskas et al. 2024; Ozyurt and Ayaz 2024; Gurcan, Soylu, and Khan 2025; Zhang et al. 2022). These methods are well suited to our analysis because employer requirements are expressed in large volumes of free text that would be difficult to examine manually at scale. Prior comparative research also suggests that transformer-based approaches such as BERTopic can produce more coherent topics than traditional methods when applied to short and unstructured text (Egger and Yu 2022). Carnevale et al. (2014) caution that job posting data can overrepresent highly skilled occupations and do not necessarily reflect actual hiring outcomes. For this reason, we interpret job postings as institutional signals rather than direct measures of employment. They provide insight into how organizations define and classify governance-related roles, as well as the skills and capabilities they value at a given time. Such signals are particularly useful in an emerging field where professional boundaries are still taking shape.

Related work provides additional support for this approach. Furnell (2021) describes an ongoing mismatch between workforce supply, employer demand, and the mix of technical and non-technical competencies required in cybersecurity. Similar patterns appear in job posting research: topic modeling of cybersecurity advertisements identifies hybrid skill expectations (Ozyurt and Ayaz 2024), while cross-country analysis shows that AI-related competencies are increasingly incorporated into existing security roles (Graham 2025). In privacy, the NIST Privacy Workforce Taxonomy offers a standardized vocabulary by linking task, knowledge, and skill statements to the NIST Privacy Framework (NIST 2024). Building on this literature, we examine employer demand for privacy professionals through large-scale job postings.

## 3. Methodology

Figure 1 summarizes the methodology as two complementary analytical phases applied to the same corpus of 1,143 job postings. The first phase focuses on data cleaning, preprocessing, and structured extraction of information such as educational requirements, certifications, regulations, competencies, and salary data. The second phase prepares the text for BERTopic analysis, including embedding, clustering, and interpretation of the resulting topics.

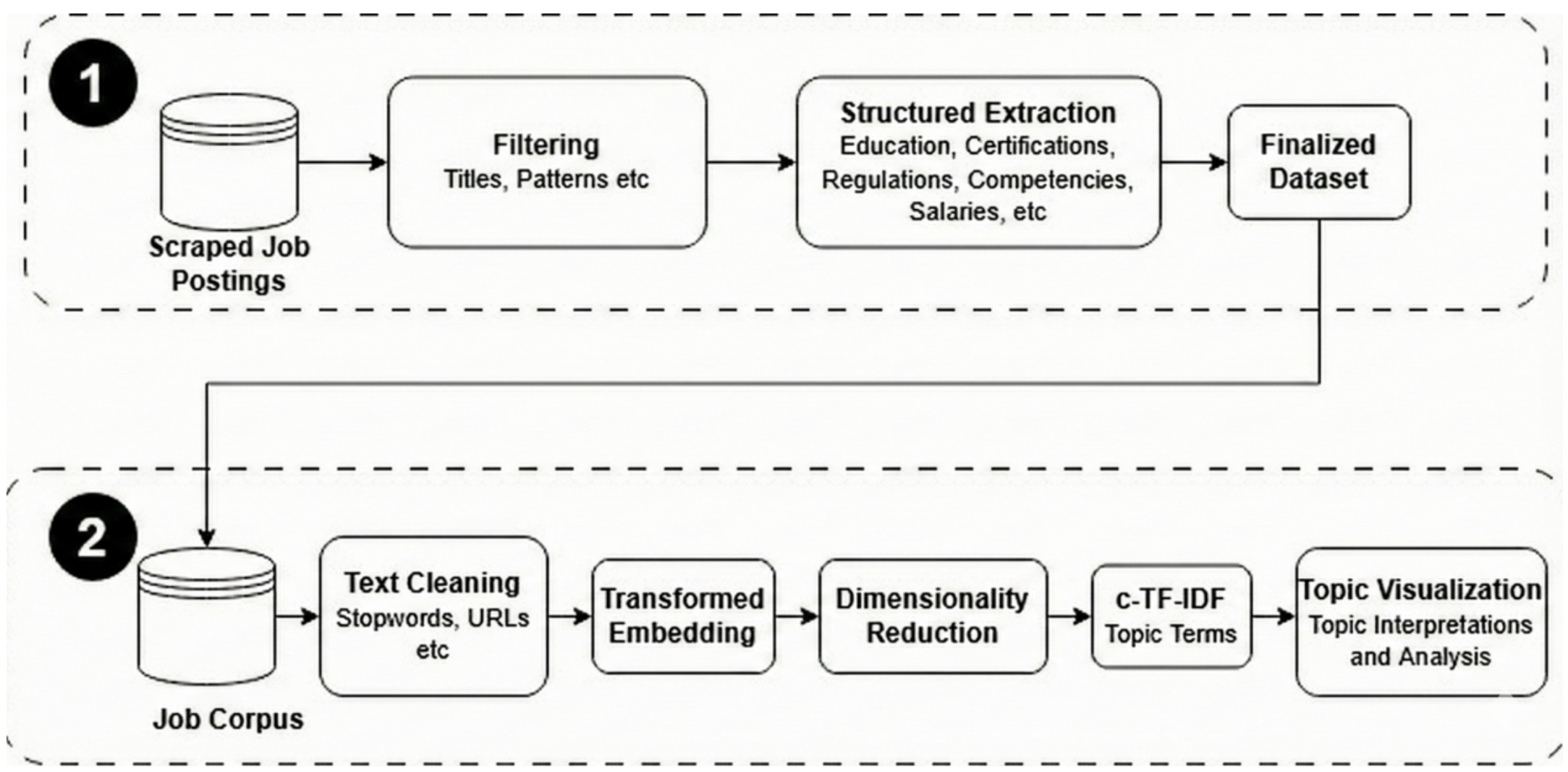

**Figure 1. Overview of the two-phase analytical workflow.**

### 3.1 Data Collection and Scope

We classify a posting as privacy-related when it is based in the United States, its title contains one of our privacy-focused search terms, and its description explicitly refers to privacy responsibilities. Using both title and description as filters helps exclude postings that mention privacy only in standard administrative language, such as equal employment opportunity statements. It also removes roles in related areas, including general security or compliance, when the job description does not identify a specific privacy responsibility. Between June and August 2025, we collected 2,254 U.S. job postings from LinkedIn and Indeed using JobSpy (Watson 2023), an open-source Python library previously used to construct large-scale job posting datasets (Kverne et al. 2025). These platforms were selected because they provide broad coverage and allow systematic data collection. We did not include Glassdoor, Dice, or ZipRecruiter because technical restrictions prevented reliable large-scale retrieval. The search terms were based on common privacy-related job titles identified in prior research (Yener, Hassan, and Bashir 2025) and on active listings from the IAPP job board during summer 2025. The complete keyword list is provided in Appendix A. Cybersecurity related job postings were retained only when their descriptions included explicit privacy-related responsibilities.

After applying the dual filter and removing postings without substantive descriptions, the final dataset included 1,143 unique job postings. All subsequent analyses are based on this filtered corpus. Because JobSpy returned unique postings, no additional cross-platform deduplication was required. We also removed postings with truncated descriptions, since incomplete text could affect the text-mining analysis. The full analysis was conducted in Python 3.10 using Google Colab, pandas, spaCy, BERTopic, and NLTK.

All job postings analyzed in this study were publicly available without login. Therefore no personal or user-generated content was collected during our data collection process. Employer names were removed from the reported results to avoid identifying individual organizations and to keep the analysis focused on broader workforce patterns.

### 3.2 Study 1: Extraction and Measurement

We extracted seven categories of variables from the postings using rule-based pattern matching in Python. The initial pattern dictionaries were built from the NIST Privacy Workforce Taxonomy (NIST 2024), the IAPP certification family, and a curated set of major data protection laws and frameworks. We then reviewed a random sample of 100 postings to identify relevant terms that were missing from the original lists and added them to the dictionaries. Matching was case-insensitive and used word-boundary anchors to reduce false positives. This manual review was intended to improve dictionary coverage rather than evaluate extraction accuracy. The reported counts therefore represent occurrences of matched patterns in employer text, not validated measures of classification accuracy.

To standardize job titles, we combined differences in seniority labels and grouped closely related title variations under the same category. For example, “Sr. Privacy Analyst” and “Senior Privacy Analyst” were coded as the same job title. Salary information was extracted from postings reporting annual pay, using the midpoint when both lower and upper bounds were given. Certifications were matched against fifteen recognized privacy and security credentials, including CIPP, CIPM, CIPT, CISSP, CISM, CDPSE, AIGP, CISA, CRISC, Security+, ISO 27001 lead implementer credentials, and HCISPP. Education and experience requirements were identified using degree-level patterns together with field-specific keywords, while years of experience were grouped into ranges. We organized competencies into two categories. Interpersonal competencies included communication, collaboration, leadership, and strategic thinking, while technical competencies covered areas such as cloud platforms, programming languages, privacy-enhancing technologies, and risk and assurance methods. To capture multi-word skills, such as “incident response,” we also applied spaCy noun-phrase extraction. References to regulations and standards were counted by name, including GDPR, HIPAA, CCPA, CPRA, GLBA, PIPEDA, NIST, ISO 27001, and SOC 2. AI-related language was identified in job descriptions through terms such as “artificial intelligence” and “machine learning,” while job titles were searched for “AI” as a standalone term.

### 3.3 Study 2: Topic Modeling with BERTopic

Job postings are semi-structured text documents in which employers describe multiple requirements and expectations in free-form language (Pejic-Bach et al. 2020). Our rule-based analysis captures explicit elements such as certifications, regulations, and job titles, while topic modeling is used to identify broader patterns in responsibilities, domains, and employer expectations. Before applying the topic model, we lowercased the text and removed punctuation, numbers, stopwords, organization names, and URLs. We also used a domain-specific stopword list to reduce company names and recruitment boilerplate, combined each posting’s title with its description, and excluded very short entries.

We considered Latent Dirichlet Allocation as a baseline approach (Blei, Ng, and Jordan 2003), but did not use it because its bag-of-words representation and fixed Dirichlet priors are less effective for short to medium-length texts with dense semantic content (Jelodar et al. 2019). Instead, we used BERTopic, which creates transformer-based sentence embeddings, reduces their dimensionality, clusters semantically similar documents, and generates interpretable topic terms through c-TF-IDF (Grootendorst 2022). Prior comparative studies report that BERTopic can match or outperform LDA on short and noisy text collections (Egger and Yu 2022; Ma et al. 2025). It has also been used to analyze large collections of job postings in related domains (Blumelhuber and Langer 2026).

We embedded each job posting using BERTopic's sentence-transformer model, selecting MiniLM for its lower computational cost and strong performance at our dataset size (Wang et al. 2020). UMAP was then used for dimensionality reduction (McInnes, Healy, and Melville 2018), followed by HDBSCAN for clustering and outlier detection (McInnes, Healy, and Astels 2017). Topic terms were generated with c-TF-IDF. The number of topics was not set in advance; the final 18 topics emerged from HDBSCAN under BERTopic's default configuration. Topic labels were assigned through consensus coding. One researcher proposed initial labels based on keywords and representative postings, while the other two reviewed them independently. All three researchers then discussed the labels and broader workforce categories until full agreement was reached.

## 4. Findings from Rule-Based Text Mining and Descriptive Analysis

### 4.1 Role Landscape

The job titles point to a diverse occupational landscape rather than a single standardized professional identity. Privacy Counsel (59) and Privacy Analyst (48) are the most common titles, followed by Privacy Officer (33), Privacy Manager (29), Privacy Specialist (29), and Privacy Engineer (28). Overall, the dataset includes 22 distinct title families, with no title appearing more than 59 times. The relatively small number of Privacy Engineer positions suggests that technical privacy work may often be incorporated into broader legal and governance roles. Data Protection Officer and Data Protection Manager titles also appear 20 times in this U.S.-only dataset, which may reflect the influence of GDPR-related governance expectations across jurisdictions. Compliance-focused titles are also common, including Compliance Analyst (27), Compliance Manager (24), and Compliance Specialist (23), suggesting that some privacy work is advertised under broader compliance labels. Table 1 presents the most common privacy job titles.

| Title | Count | Includes |
|---|---|---|
| Privacy Counsel | 59 | Legal/Corporate Privacy Counsel, Sr Privacy Counsel, Global Privacy Counsel |
| Privacy Analyst | 48 | Sr. Privacy Analyst, Senior Privacy Analyst, Data Privacy Analyst, Technical Privacy Analyst, etc. |
| Privacy Officer | 33 | Chief Privacy Officer, VP Compliance & CPO, Sr. Privacy Officer, Privacy Officer |
| Privacy Manager | 29 | Data Privacy Manager, Privacy and Data Protection Manager, Privacy Compliance Manager |
| Privacy Specialist | 29 | Senior/Associate Privacy Specialist, Privacy & Data Protection Specialist |
| Privacy Engineer | 28 | Sr. Privacy Engineer, Privacy Engineer, Privacy Platform Engineer, Principal Privacy Engineer |
| Compliance Analyst | 27 | Payment Compliance Analyst, Risk & Compliance Analyst |
| Privacy Program Manager | 14 | Program Manager, Privacy & Trust; Privacy Operations Manager |
| Compliance Manager | 24 | Sr. IT Cybersecurity & Compliance Manager, Regulatory Compliance Manager |
| Compliance Specialist | 23 | Corporate Compliance Specialist, Clinical Compliance Specialist |
| Privacy Attorney / Lawyer | 23 | Privacy Attorney, Privacy Lawyer, Counsel, Privacy & AI Governance |
| Compliance Officer | 22 | Tobacco Compliance Officer, Compliance & Privacy Officer, Trust Compliance Officer |
| Data Protection Officer / Manager | 20 | DPO, Data Protection Manager, Data Protection Analyst |
| Compliance Director / Associate Director | 20 | Director, US Compliance, Director of Compliance & Risk Advisory |
| Data Governance Manager / Specialist | 19 | Data Governance Analyst, Director of Data Governance |
| Governance, Risk & Compliance (GRC) Analyst / Manager | 19 | Governance, Risk & Compliance (GRC) Analyst / Manager |
| Privacy Coordinator / Associate / Fellow | 18 | Privacy Coordinator, Privacy Associate, Privacy Fellow |
| Privacy Director | 18 | Director of Privacy & Senior Legal Counsel, Director, Privacy by Design |
| Compliance Program Manager / Lead | 17 | Compliance Program Manager -- Privacy, Compliance Program Lead |
| Privacy Consultant | 16 | Lead Privacy Consultant, Senior Privacy Consultant |
| Privacy Awareness / Risk / Enablement Specialist | 16 | Privacy Awareness / Risk / Enablement Specialist |
| Technical Product Manager, Privacy (AI / Product) | 16 | Privacy Technical Product Manager |

**Table 1. Privacy job titles.**

### 4.2 Salary Distribution

Salary disclosures vary considerably across postings. The mean annual salary is $146,467 and the median is $135,425, with reported values ranging from $29,120 to $371,000. The higher mean reflects a right-skewed distribution, with a small number of highly paid roles raising the average. Among cities with at least five postings, Mountain View, CA has the highest mean salary at $239,258 and a median of $221,600, followed by San Francisco, CA ($216,881) and Los Angeles, CA ($198,272). These figures suggest that privacy roles linked to technology sectors may offer higher compensation in high-cost metropolitan areas. Figure 2 shows the salary distribution, while Table 2 presents mean and median salaries by city.

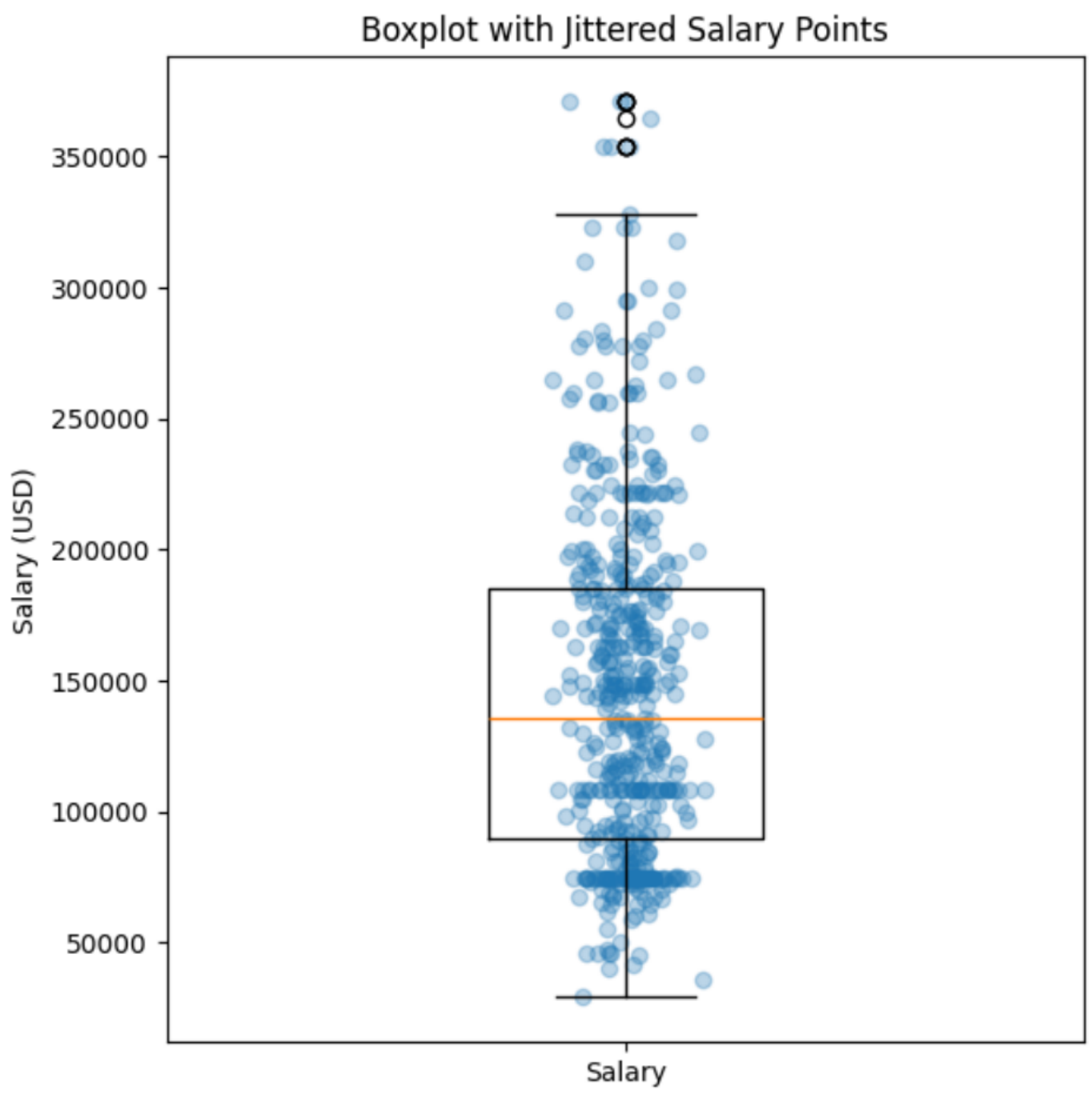

**Figure 2. Salary distribution (boxplot with jittered points) for postings that disclose compensation.**

Salary findings should be interpreted with caution because postings that report compensation do not represent a random subset of the corpus. Disclosure requirements differ across U.S. jurisdictions, and reporting practices also vary by employer size and sector. As a result, postings with salary information are more likely to come from places and organizations where disclosure is required or common, including several high-cost metropolitan areas. These figures therefore describe only the disclosing subset and may overstate the central tendency of salaries across the broader privacy workforce.

### 4.3 Certifications

Certifications are a strong signal in the dataset. The Certified Information Privacy Professional (CIPP) is mentioned most often (373), followed by the Certified Information Privacy Manager (CIPM) (251) and the Certified Information Privacy Technologist (CIPT) (169). Security-focused credentials are also common, including the Certified Information Systems Security Professional (CISSP) (145), Certified Information Security Manager (CISM) (70), and Certified Data Privacy Solutions Engineer (CDPSE) (34). The prominence of both privacy-specific and broader security and governance credentials suggests that employers position privacy expertise within a wider security, risk, and compliance ecosystem. The Artificial Intelligence Governance Professional (AIGP) appears 16 times, further indicating a growing connection between AI oversight and privacy practice. Figure 3 presents the most frequently requested certifications.

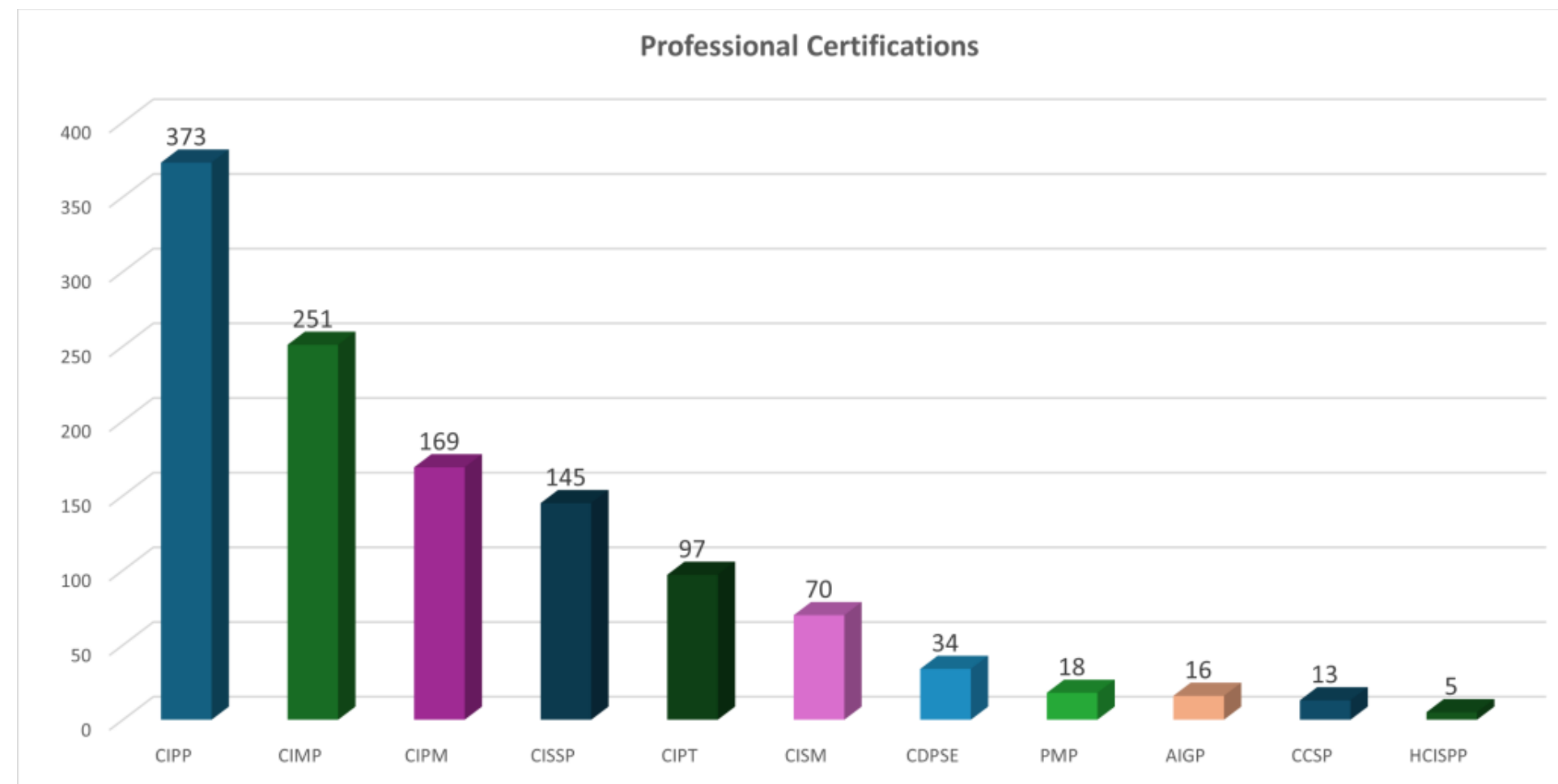

**Figure 3. Most frequently requested certifications in privacy-related postings.**

### 4.4 Education and Experience

Privacy roles draw from a wide range of educational backgrounds. Degree requirements appear in 73 percent of postings, including bachelor's degrees (588 mentions), master's degrees (199), PhDs (41), and high school qualifications (12). Computer Science and related engineering fields are mentioned most often (227), followed by Mathematics and Quantitative Fields (117) and Information Systems (107). STEM and computing categories account for 682 mentions. Business and Management (256) and Law and Policy (119) also appear frequently, reinforcing the view that privacy roles span both technical and governance domains. Table 3 presents the full distribution.

Most experience requirements are concentrated at the mid-career level. Thirty-five percent of postings ask for four to six years of experience, while 29 percent seek one to three years. Another 18 percent do not specify an experience requirement, which may reflect entry-level opportunities or greater flexibility in hiring criteria.

### 4.5 Competencies

The competency patterns further illustrate the sociotechnical nature of privacy work in our dataset. The most frequently mentioned competencies are interpersonal, including communication (1,030 mentions), collaboration (856), leadership (784), and analytical thinking. Each appears more often than any individual technical competency.

Technical competencies are also common across the postings. Cloud technologies and platforms, including AWS, Azure, and Google Cloud Platform, appear in 599 postings. Cloud platforms, encryption, and privacy impact assessment are each mentioned more than one hundred times, while privacy impact assessment appears in 107 postings and privacy-enhancing technologies in 35. Programming languages are less frequent, with Python mentioned in 96 postings and SQL in 66. The high frequency of cloud-related terms suggests that cloud infrastructure is becoming closely connected to privacy functions.

These patterns suggest that employers expect privacy professionals to operate across both organizational and technical domains rather than specialize in only one area. Figure 4 presents the most frequently mentioned technical competencies, while Figure 5 shows the most common interpersonal competencies.

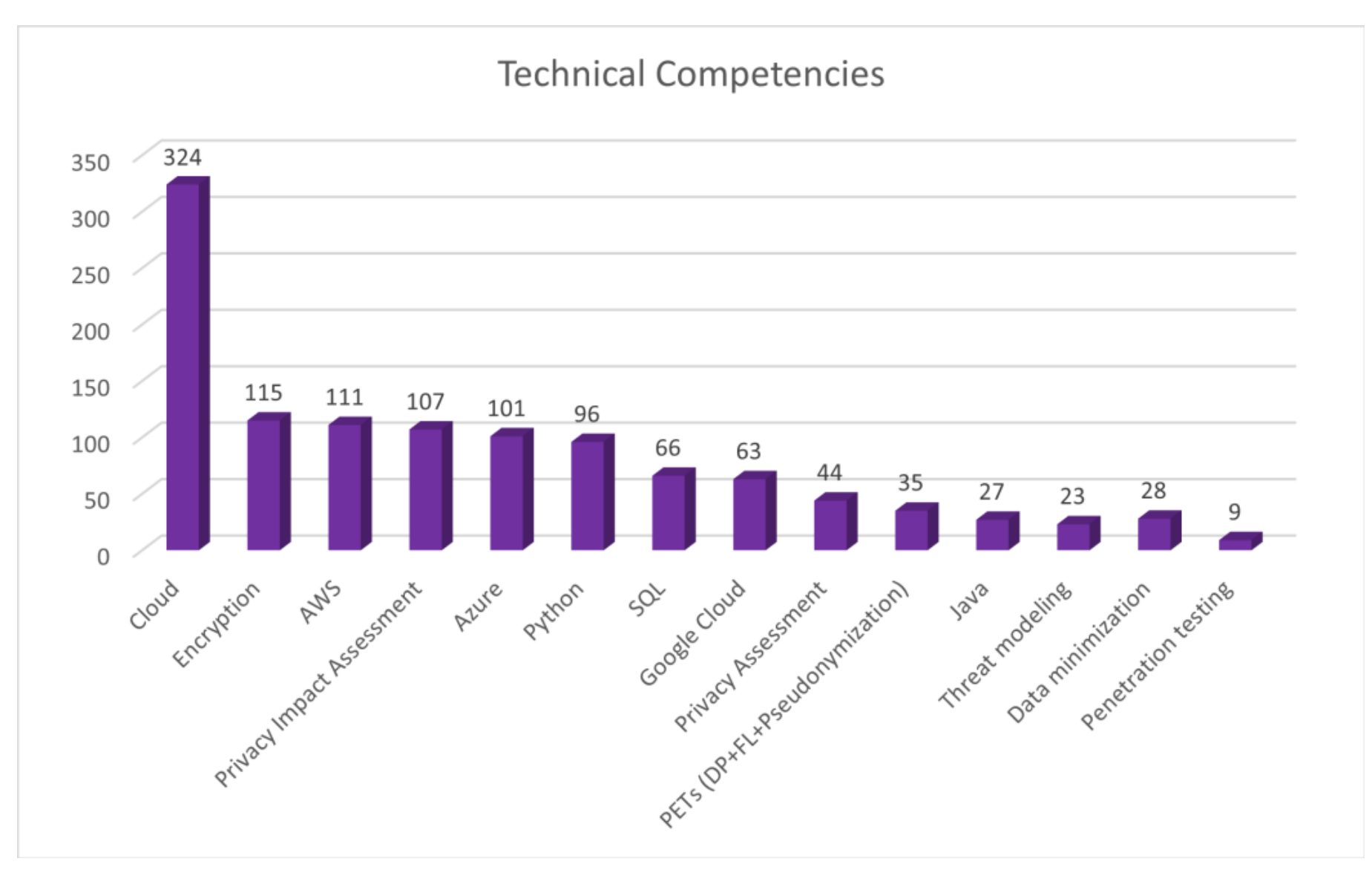

**Figure 4. Technical competencies in privacy job postings.**

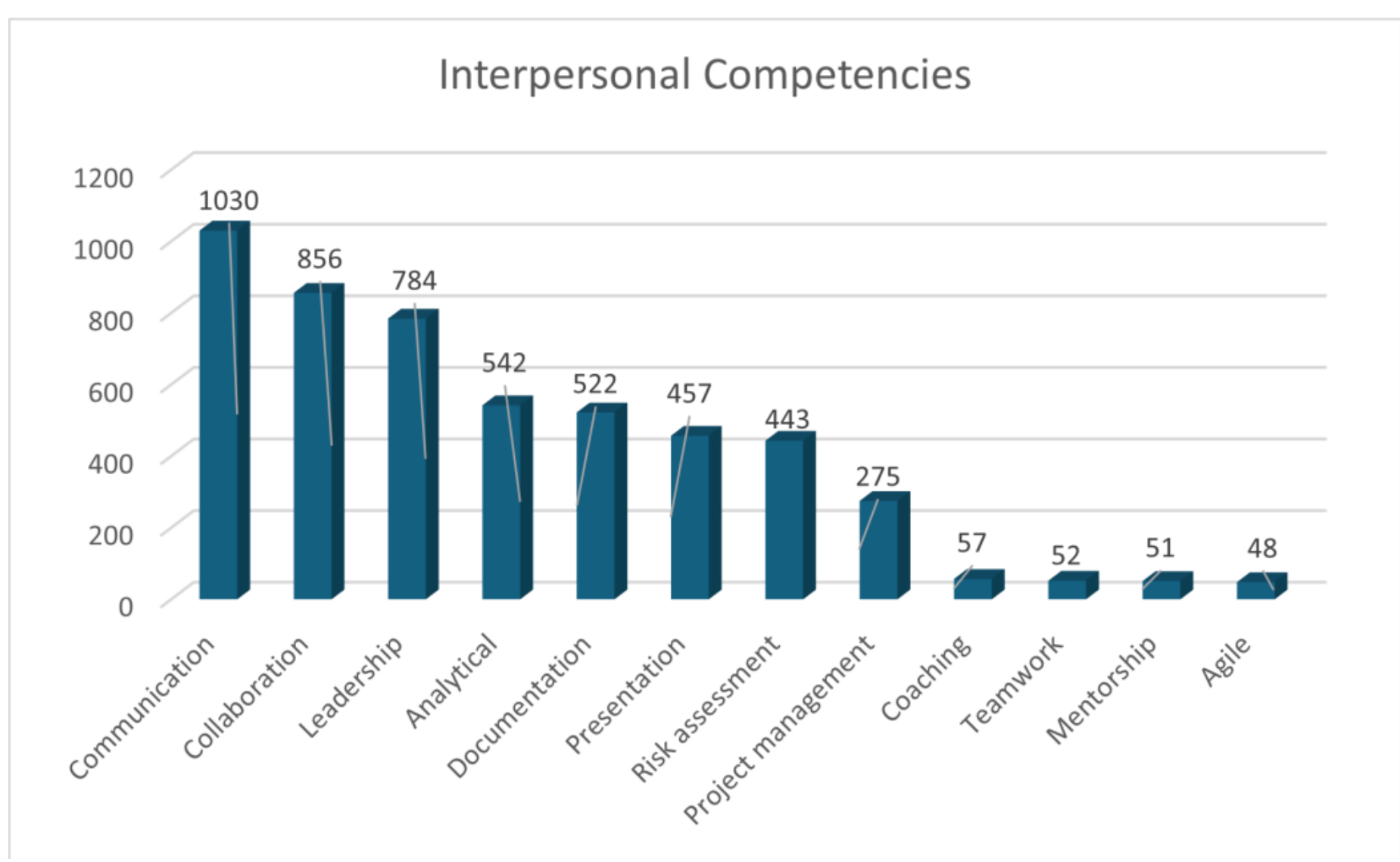

**Figure 5. Interpersonal competencies in privacy job postings.**

### 4.6 Regulations and Standards

Regulatory references also reflect an international orientation. GDPR is mentioned most often (594), followed by HIPAA (525), CCPA (472), and CPRA (164). Given that the dataset includes only U.S. postings, this pattern suggests that employers view regulatory literacy as extending beyond a single jurisdiction (European Parliament and Council 2016).

References to frameworks also point to a broader risk-management focus. NIST appears 322 times, followed by ISO 27001 (124) and SOC 2 (80), placing privacy roles within wider enterprise governance structures (NIST 2020). Figure 6 presents the most frequently mentioned regulations in the postings.

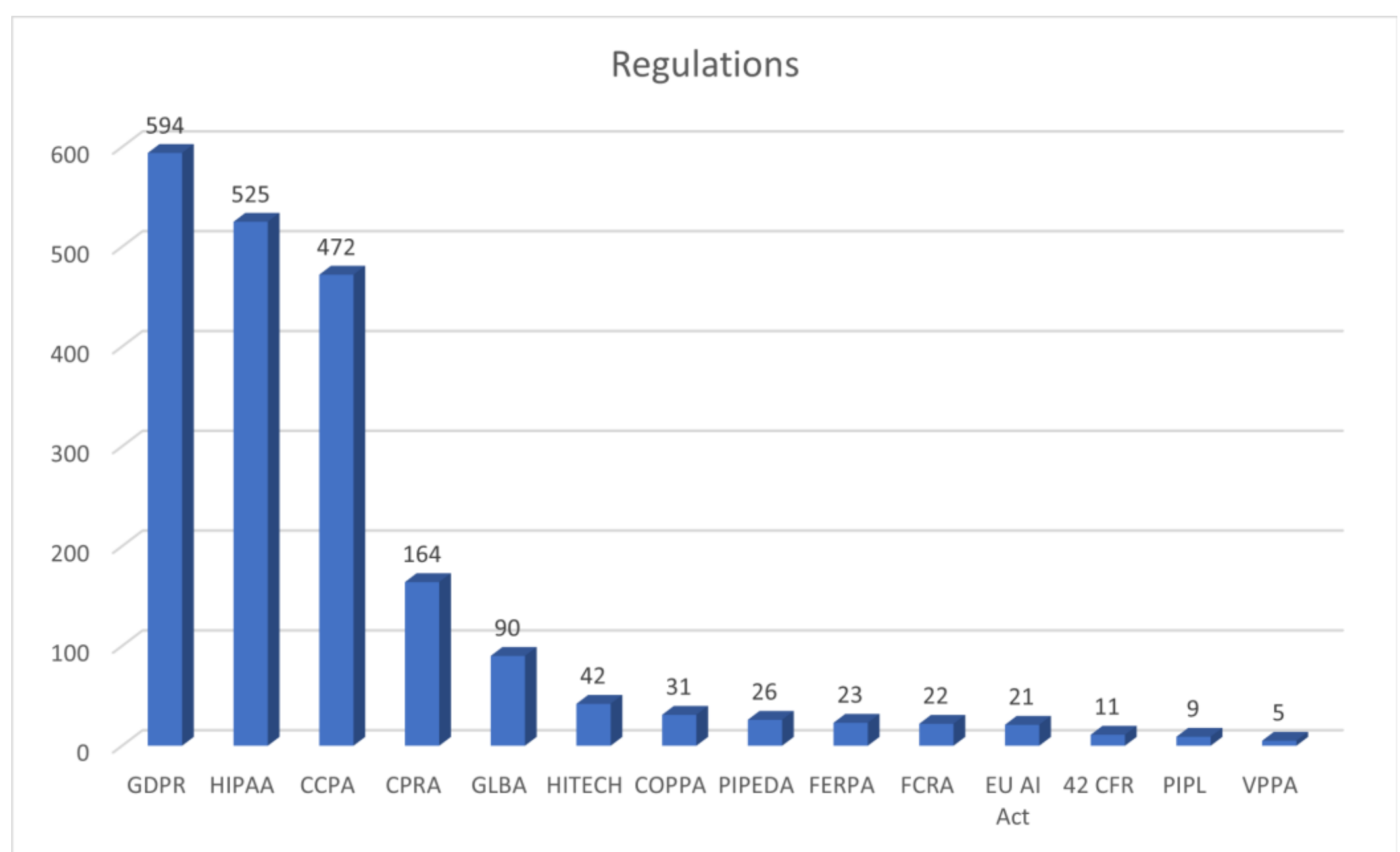

**Figure 6. Regulations in privacy job postings.**

### 4.7 AI Mentions

The term "artificial intelligence" appears in 51 percent of the postings, and 68 job titles explicitly include "AI." In several cases, AI responsibilities are added to roles that are not exclusively focused on AI. For example, "Global Privacy Counsel" appears as "Global Privacy Counsel, AI Governance," while "VP, Product Privacy" appears as "VP, Product Privacy and AI Legal." This pattern suggests that AI governance responsibilities are often incorporated into existing privacy positions and is consistent with survey evidence showing that privacy leaders are increasingly responsible for AI oversight (IAPP 2024a).

## 5. Study 2 Findings: Latent Themes from Topic Modeling

Our BERTopic analysis identified 18 topics representing different thematic areas of the privacy workforce. Topics 0 and 1 are the largest, together accounting for 38.4 percent of all postings. Topic 0 is characterized by terms such as “compliance,” “security,” “risk,” and “management,” while Topic 1 includes frequent terms such as “compliance,” “privacy,” “health,” and “healthcare.” Figure 7 presents the topic word scores for these leading topics.

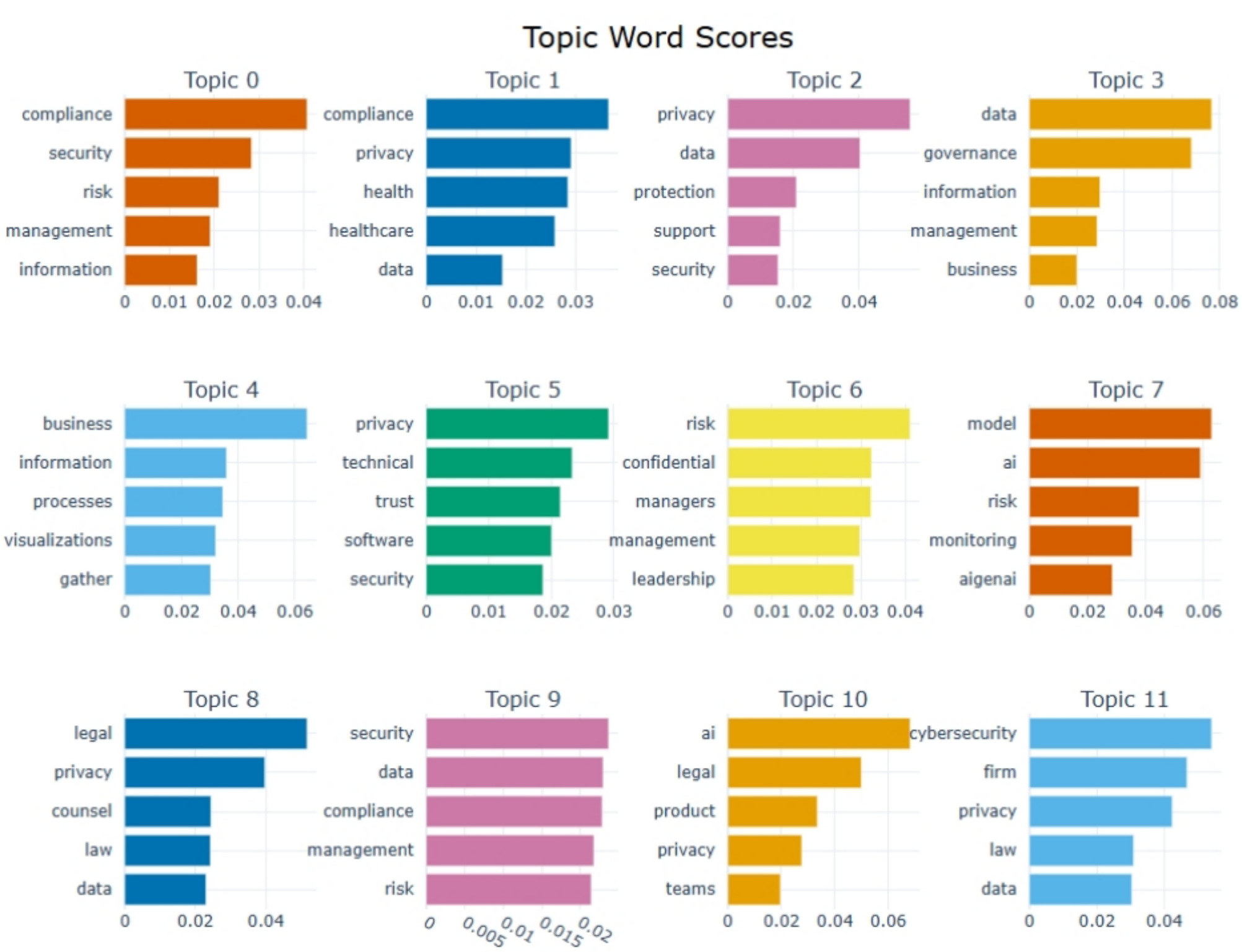

Figure 7. Topic word scores for the most dominant topics. Bar length gives the c-TF-IDF weight of each term within its topic.

To support topic labeling and grouping, we created the inter-topic similarity matrix shown in Figure 8. After reviewing the relationships among topics, three researchers agreed on four broader categories: Compliance and Risk, Governance and Management, Legal and Regulatory, and Engineering. Table 4 presents each topic together with its keywords, label, and category. Some topics within the Legal and Regulatory category share similar keywords. Topics 8, 12, and 14, for example, all involve legal counsel and compliance but differ in their specific focus. A broader clustering approach might have combined them into one legal compliance theme, but we kept them separate because their representative postings reflect different areas of legal responsibility. Even so, the boundaries among these topics are less distinct than those between more clearly separated topics such as Topic 5 and Topic 1.

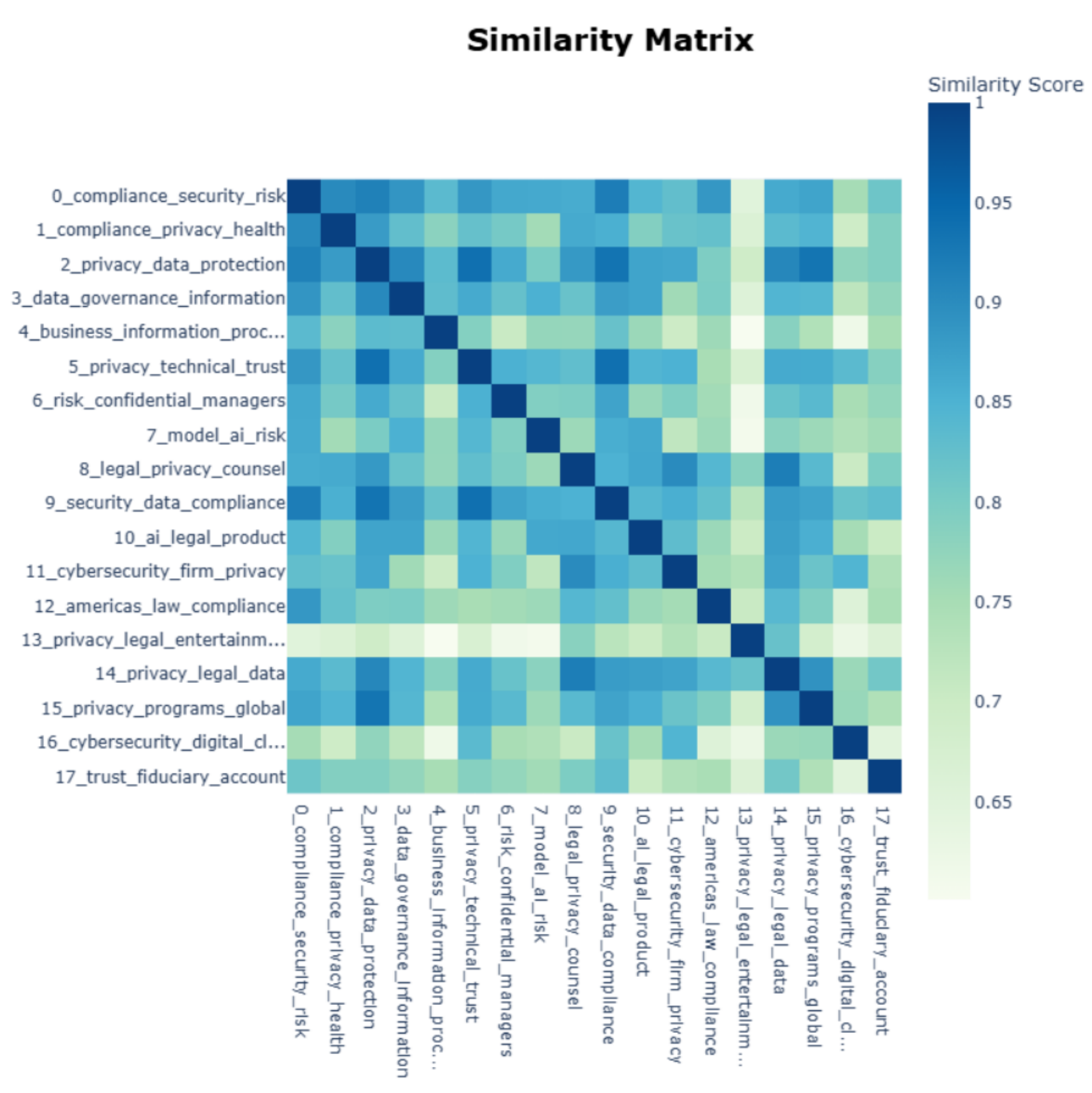

**Figure 8. Inter-topic similarity matrix.**

| Topic # | Percentage | Keywords | Labels | Main Category |
|---|---|---|---|---|
| 0 | 20.21% | compliance, security, risk, management, information, business, regulatory, internal, privacy, support | Security Compliance & Risk Management | Compliance & Risk |
| 1 | 18.20% | compliance, privacy, health, healthcare, data, regulatory, hipaa, regulations, laws, information | Healthcare Privacy Compliance | Compliance & Risk |
| 6 | 4.46% | risk, confidential, managers, management, leadership, confidentiality, teams, identify, risks, deliverables | Privacy Risk Management and Leadership | Compliance & Risk |
| 7 | 3.85% | model, ai, risk, monitoring, aigenai, consulting, officers, clients, help, deployment | AI/ML Risk Monitoring | Compliance & Risk |
| 9 | 2.10% | security, data, compliance, management, risk, privacy, business, access, wikimedia, clients | Data Security Compliance | Compliance & Risk |
| 2 | 9.80% | privacy, data, protection, support, security, information, teams, global, business, management | Global Privacy Management | Governance & Management |
| 3 | 4.90% | data, governance, information, management, business, policies, support, tools, standards, ai | Data Governance and Management | Governance & Management |
| 4 | 4.72% | business, information, processes, visualizations, gather, data, dashboards, develops, detailed, discovery | Information and Business Process | Governance & Management |
| 15 | 1.14% | privacy, programs, global, business, impact, professionals, organizations, president, chief, comprehensive | Global Privacy Program Leadership | Governance & Management |
| 17 | 0.96% | trust, fiduciary, account, insurance, trusts, client, estate, policy, needs, asset | Data Stewardship | Governance & Management |
| 8 | 3.76% | legal, privacy, counsel, law, data, business, laws, compliance, support, litigation | Legal Compliance | Legal & Regulatory |
| 10 | 1.75% | ai, legal, product, privacy, teams, global, governance, data, law, laws | AI Legal Governance | Legal & Regulatory |
| 11 | 1.49% | cybersecurity, firm, privacy, law, data, practice, clients, litigation, breach, response | Legal and Cybersecurity Practice | Legal & Regulatory |
| 12 | 1.31% | americas, law, compliance, legal, counsel, global, business, diverse, support, regional | Regional & Global Legal Compliance | Legal & Regulatory |
| 13 | 1.31% | privacy, legal, entertainment, data, audio, advertising, technology, business, ad, trade | Entertainment Privacy & Advertising Law | Legal & Regulatory |
| 14 | 1.31% | privacy, legal, data, protection, product, business, teams, global, counsel, law | Business Data Protection & Legal Compliance | Legal & Regulatory |
| 5 | 4.55% | privacy, technical, trust, software, security, data, client, engineering, technology, management | Privacy Engineering | Engineering |
| 16 | 1.05% | cybersecurity, digital, clients, organizations, business, data, challenges, consulting, teams, developing | Organizational Cybersecurity Challenges | Engineering |

**Table 4. BERTopic labels based on keyword-derived topics (four-category structure).**

## 5.1 Compliance and Risk

This category includes Topics 0, 1, 6, 7, and 9 and reflects a continued emphasis on regulatory compliance and risk mitigation. Security Compliance and Risk Management (Topic 0) is the largest topic, accounting for 20.2 percent of postings and combining compliance, security, and risk management responsibilities. Healthcare Privacy Compliance (Topic 1), at 18.2 percent, centers on health data regulation and HIPAA.

Privacy Risk Management and Leadership (Topic 6), at 4.5 percent, includes risk identification, protection of confidential data, and team management. AI/ML Risk Monitoring (Topic 7), at 3.8 percent, focuses on monitoring AI models for regulatory compliance and advising privacy teams on AI ethics. Data Security Compliance (Topic 9), at 2.1 percent, links privacy with security operations through responsibilities such as access control enforcement and collaboration between privacy and security teams.

### 5.2 Governance and Management

Topics 2, 3, 4, 15, and 17 established this category, and reflects growing expectations for strategic oversight and long-term stewardship of personal information.

Global Privacy Management (Topic 2), at 9.8 percent, covers international privacy programs, cross-border collaboration, and consistent privacy controls across global teams. Data Governance and Management (Topic 3), at 4.9 percent, covers policy design, AI governance, and maintenance of privacy standards. Information and Business Process (Topic 4), at 4.7 percent, covers documentation, business information flows, data gathering, and communication of insights through dashboards and visualizations. Global Privacy Program Leadership (Topic 15), at 1.1 percent, reflects senior roles such as Chief Privacy Officer that build comprehensive global programs and raise staff privacy awareness. Data Stewardship (Topic 17), at 1.0 percent, points to the fiduciary responsibilities of privacy professionals and to ethical governance of client data.

### 5.3 Legal and Regulatory

This category includes Topics 8, 10, 11, 12, 13, and 14, and shows demand for legal expertise spanning AI regulation and domain-specific data governance.

Legal Compliance (Topic 8), at 3.8 percent, connects legal counsel with compliance oversight, advising on privacy laws and interpreting regulations. AI Legal Governance (Topic 10), at 1.7 percent, reflects legal oversight of AI products and emerging regulatory concerns around AI governance frameworks. Legal and Cybersecurity Practice (Topic 11), at 1.5 percent, covers law-firm work on breach response and incident-related legal obligations. Regional and Global Legal Compliance (Topic 12), at 1.3 percent, covers counsel roles managing compliance across regions and supporting regional business units. Entertainment Privacy and Advertising Law (Topic 13), at 1.3 percent, covers advertising compliance and digital media regulation, including data used in content, audio, and targeted advertising. Business Data Protection and Legal Compliance (Topic 14), at 1.3 percent, covers legal support for data protection in product development and business operations.

### 5.4 Engineering

This category includes Topics 5 and 16 and captures the technical expectations placed on privacy professionals.

Privacy Engineering (Topic 5), at 4.5 percent, reflects technical implementation of privacy and security features in software systems, where privacy engineers embed controls, collaborate with development teams, support cybersecurity needs, and address technical risks created by digital transformation. Organizational Cybersecurity Challenges (Topic 16), at 1.0 percent, centers on developing cybersecurity solutions and consulting on digital security challenges, connecting privacy, security, and engineering.

Overall, the topic structure shows a privacy workforce characterized by multidisciplinary specialization across all four categories. Table 4 lists the 18 topics in detail.

# 6. Discussion

## 6.1 What the Two Studies Show Together

The two analytical approaches offer complementary views of the same 1,143 job postings. Rule-based text mining captures requirements that employers state explicitly, while BERTopic reveals how those requirements cluster across the corpus. Considered together, the results portray a privacy workforce that spans traditional professional boundaries. This pattern is especially clear in the competencies employers request. Communication, collaboration, leadership, and analytical skills are among the most common requirements, while technical expertise also remains important. Cloud platforms, encryption, and privacy impact assessment each appear in more than one hundred postings. The topic model shows a similar combination: risk management appears alongside leadership, and technical implementation is linked with client and management responsibilities. The largest topic also brings together compliance, security, risk, management, and business language. In employer descriptions, privacy work therefore does not divide neatly into technical and non-technical roles.

The diversity of job titles points to the same pattern. We identified 22 title families, and no single title appeared more than 59 times. Although Privacy Counsel was the most common, it still accounted for only a small portion of the corpus. This variation suggests that privacy is not represented as one standardized occupation. Instead, privacy responsibilities are distributed across legal, technical, governance, risk, and managerial roles.

AI governance is an important part of this pattern. AI-related language appears in 51 percent of the postings, and 68 job titles explicitly mention AI. However, these references are spread across several areas of the topic model rather than concentrated in one cluster. Topic 7 links AI and machine learning with risk monitoring and compliance, Topic 10 places AI within legal governance, and Topic 3 connects it with data governance and management. This suggests that AI governance is being incorporated into multiple forms of privacy work rather than appearing only in distinct AI-focused positions. However, our data cannot show whether a separate AI governance occupation is emerging elsewhere in the labor market, because the sampling strategy was not designed to answer that question.

The connection between privacy and security is also clear in the dataset. CISSP is mentioned in 145 postings and CISM in 70, while cloud technologies appear frequently across roles. Security Compliance and Risk Management is also the largest topic identified by the model, and other topics connect privacy with access control, cybersecurity, and breach response. These patterns suggest that many privacy positions sit at the intersection of regulatory obligations, organizational risk, and security controls. This relationship is consistent with Radanliev's (2024) discussion of the boundary between regulation and technical security.

The two methods are not expected to produce identical findings because they capture different aspects of the postings. Some competencies identified through rule-based extraction, for example, do not appear as distinctive terms in the topic model. Rule-based extraction detects specific requirements wherever they occur, while topic modeling highlights language that differentiates groups of postings. The value of using both methods therefore comes from their complementary functions rather than from producing the same results.

**6.2 Privacy Roles as Hybrid Governance Positions**

Prior research describes privacy professionals as helping connect legal requirements with technical implementation (Samarin et al. 2025). Our findings suggest that this role now extends beyond that relationship. Employers also expect privacy professionals to communicate across teams, manage risk, coordinate organizational processes, and, in many cases, assume responsibilities related to AI governance.

From this perspective, hybridity involves more than expecting one employee to combine legal and technical expertise. Privacy responsibilities may be distributed across several organizational roles, while individual positions still bring together duties that are often treated as separate specialties. This aligns with the broader concept of boundary-spanning work, which can be shared across multiple actors rather than carried out by a single individual (Friedman and Podolny 1992).

Job postings cannot show how these expectations translate into actual work after hiring. Employers may identify candidates who match the full profile, revise their expectations during recruitment, or describe an ideal candidate who is difficult to find. What the postings reveal is how organizations currently define the work they expect privacy professionals to perform.

### 6.3 Workforce Identity and the Title Mismatch

Another issue is how privacy work is labeled in the labor market. Although privacy engineering is often treated as a distinct specialization, the titles in our corpus are much more varied. Legal, analytical, governance, managerial, and engineering roles all appear, with Privacy Counsel as the most common title.

Technical privacy responsibilities are not confined to roles titled Privacy Engineer. Postings with other titles also mention privacy impact assessments, cloud technologies, and reviews of technical controls. The topic model shows a similar pattern, with technical language appearing across compliance, governance, and legal topics in addition to the Engineering category. This suggests that technical privacy work extends beyond positions explicitly labeled as engineering roles. This has an important methodological implication. Studies that define the privacy workforce using job titles alone may overlook technical privacy work listed under legal, compliance, or governance roles.

### 6.4 Institutional and Educational Implications

The postings reflect a regulatory environment that extends beyond U.S. law. Although the corpus includes only U.S. positions, GDPR is the most frequently cited regulation, while references to NIST and ISO are also common. Employers therefore describe privacy work through a combination of domestic law, international regulation, security frameworks, and enterprise risk practices. This combination also helps explain the frequent connection between privacy and security responsibilities in the corpus. About one quarter of the postings fall within topics related to security compliance, security operations, cybersecurity, or breach response. Privacy roles therefore often require competencies and credentials that are also common in security-focused work.

AI adds another dimension to this organizational connection. Responsibilities involving AI risk, data protection, legal compliance, and system oversight appear across several types of privacy positions. A key question for future research is how organizations distribute these responsibilities in practice and where accountability is ultimately placed. Job postings cannot resolve that issue, but they indicate why it warrants further study.

Education presents a related challenge. Employers describe privacy roles as requiring a combination of legal, technical, and interpersonal knowledge rather than expertise in a single area. Programs that focus only on one of these domains may therefore address only part of the profile employers seek. Interdisciplinary privacy programs may offer one response, but assessing how well existing curricula align with labor-market expectations would require separate analysis.

## 7. Limitations

The dataset captures only a specific segment of the U.S. privacy labor market and should not be treated as a complete representation of privacy employment. We collected postings from LinkedIn and Indeed over a three-month period in summer 2025 through a single retrospective collection. As a result, positions that had already been filled, withdrawn, or expired before collection are not included. Because we do not have records of those missing postings, we cannot assess how their absence may have influenced the observed distribution of roles.

The sampling strategy also limits what we can conclude about AI governance. We searched for privacy-related job titles and required each description to include privacy-related language. This design was appropriate for examining how AI governance appears within the privacy workforce, but it excludes AI governance roles advertised separately from privacy. The findings therefore should not be interpreted as evidence of the overall size or development of the AI governance workforce. Examining whether distinct AI governance occupations are emerging alongside privacy roles would require a broader sampling approach.

Job postings themselves offer only one perspective on occupational practice. They capture how employers describe positions during recruitment rather than the responsibilities employees ultimately perform or the qualifications of those who are actually hired (Carnevale, Jayasundera, and Repnikov 2014; Dawson et al. 2019). Some requirements may also reflect standardized or aspirational language. This limitation is especially relevant for frequently mentioned interpersonal competencies such as communication and collaboration.

AI-related language should be interpreted with similar caution. References to AI may be increasingly common across professional job advertisements in general rather than distinctive to privacy work. Because our study does not include a comparison corpus of non-privacy positions, we cannot determine whether this pattern is unique to privacy, although related research has also identified AI competencies in cybersecurity roles (Graham 2025). Our findings therefore show that AI responsibilities are widespread within privacy hiring, but not that they are specific to the privacy workforce.

## 8. Future Work

Several extensions could build on these findings. A longitudinal study would help determine whether the patterns observed here reflect an ongoing shift in the privacy profession. Repeated data collection over a longer period could show whether AI-related responsibilities continue to increase, whether new job titles emerge, and whether privacy roles become more specialized or remain closely connected to security, legal, and governance functions.

Comparative data could also help clarify the significance of the AI findings. A matched corpus of non-privacy professional postings could show whether the prevalence of AI-related language is distinctive to privacy or part of a broader shift in professional hiring. Cross-jurisdictional comparisons could further examine whether similar occupational patterns appear in other regulatory environments, particularly in the European Union and the United Kingdom, where AI-specific regulation may shape hiring differently.

Future research could directly compare employer expectations in job postings with evidence from privacy professionals. Prior interview and survey research has examined the responsibilities and experiences of privacy practitioners, while our study captures how employers describe these roles during recruitment. Comparing these perspectives could show where advertised expectations align with or differ from the responsibilities professionals report performing in practice, particularly across legal, technical, managerial, and AI governance functions.

## 9. Conclusion

This study examined how U.S. employers define the privacy workforce through an analysis of 1,143 job postings. Using rule-based text mining and BERTopic, we identified a diverse occupational landscape spanning legal, technical, governance, risk, and managerial functions. Employers often seek a combination of technical knowledge, regulatory expertise, and interpersonal skills, suggesting that privacy work does not fit neatly within a single professional domain.

AI governance is also embedded throughout this landscape. AI-related language appears in 51 percent of the postings and across compliance, legal, and governance topics rather than forming a distinct occupational cluster. This suggests that organizations are incorporating AI governance into existing privacy functions alongside more established responsibilities for data protection, compliance, risk management, and security. At the same time, the variety of job titles and responsibilities indicates that this work is distributed across several types of positions rather than organized around a single professional identity.

Overall, the findings position privacy work as a multidisciplinary area of organizational governance in which legal, technical, security, and managerial responsibilities increasingly intersect with AI-related oversight. This study provides an empirical view of how employers currently describe these roles in the U.S. labor market and offers a foundation for future research on how privacy and AI governance occupations develop over time.

## Acknowledgments

We gratefully acknowledge Alyssa Rachel Kalish for her contribution to data collection from job platforms.

## Appendices

### A. Search Keywords

For completeness, we provide the exact keywords used to collect the job postings. Each term was submitted to the JobSpy scraper with the location set to the United States. The retrieved postings were then screened using the dual title-and-description filter described in Section 3.1.

“privacy engineer; privacy analyst; privacy compliance; privacy officer; data privacy; data protection; governance and information officer; privacy lawyer; privacy program management; director, privacy; data privacy analyst; privacy coordinator; privacy and data security; information governance risk analyst; program manager data protection and privacy; privacy project manager; data privacy protection; privacy officer; data risk and incident manager; analyst, privacy and cybersecurity law; data strategy, privacy and security associate; access and privacy practice advisor; risk and compliance; associate general counsel, privacy; manager freedom of information and privacy; privacy and security director; privacy consultant; privacy and trust engineer; data privacy and ai governance.”

**B. Tables**

## Average and median salaries by city.

| City | Mean | Median |
|---|---|---|
| Mountain View | 239,258.3 | 221,600 |
| San Francisco | 216,881.3889 | 223,300 |
| Los Angeles | 198,271.5909 | 180,500 |
| San Jose | 191,566.2778 | 194,300 |
| San Diego | 187,450 | 193,500 |
| Boston | 186,271.5 | 183,212.5 |
| New York | 177,819.9545 | 167,500 |
| Baltimore | 174,062.5 | 180,000 |
| Seattle | 173,153.7333 | 170,000 |
| Chicago | 166,067.1429 | 181,500 |
| Remote | 163,200.925 | 152,150 |
| Austin | 155,061.9118 | 164,911 |
| Washington | 135,565.7692 | 132,000 |
| Denver | 123,450 | 108,750 |
| Minneapolis | 122,571.8 | 121,649 |
| Dallas | 120,231.8333 | 108,750 |
| Atlanta | 119,213.8571 | 114,713 |
| Houston/Aton | 113,179.75 | 104,551 |
| Phoenix | 110,438.5 | 91,283 |
| Kansas City | 93,842.2 | 108,750 |

**Table 2. Average and median salaries by city, for cities with at least five postings.**

| Main Category | Degree Name | Count |
|---|---|---|
| STEM and Computing | Computer Science (CE,CS,SE) | 227 |
| | Information Technology | 64 |
| | Information Systems | 107 |
| | Information Management | 21 |
| | Information/Data Science | 25 |
| | Mathematics and Quantitative Fields | 117 |
| | Statistics | 65 |
| | Engineering (general) | 33 |
| | Data Architecture | 6 |
| | Data Management | 6 |
| | Natural Sciences | 11 |
| Business and Management | Business/Business Analytics | 78 |
| | Business Administration | 39 |
| | Business Management | 22 |
| | Economics | 62 |
| | Finance | 36 |
| | Accounting | 19 |
| Law and Policy | Law | 74 |
| | Juris Doctor | 6 |
| | Legal Studies | 11 |
| | Public Policy | 16 |
| | Policy | 3 |
| | Public Administration | 4 |
| | Criminal Justice | 2 |
| | Political Science | 2 |
| | International Relations | 1 |

| Main Category | Degree Name | Count |
|---|---|---|
| Health and Healthcare | Healthcare Administration | 16 |
| | Health Care Administration | 6 |
| | Public Health | 10 |
| | Health Information Management | 5 |
| | Health Information Technology | 1 |
| | Health Administration | 1 |
| | Nursing | 3 |
| | Health Informatics | 3 |
| | Clinical Fields | 2 |
| Social Sciences and Humanities | Psychology/Sociology/Social Work | 55 |
| | Communications | 7 |
| | Library Science | 2 |
| | Arts | 2 |
| Privacy Governance and Risk | Privacy Engineering | 13 |
| | Data Privacy | 3 |
| | Data Governance | 9 |
| | Governance (IT Governance) | 5 |
| | Risk Management | 11 |
| | Compliance | 13 |

**Table 3. Degree expectations for the privacy workforce.**

## References

Bamberger, Kenneth A., and Deirdre K. Mulligan. 2011. "Privacy on the Books and on the Ground." Stanford Law Review 63 (2): 247-315.

Blei, David M., Andrew Y. Ng, and Michael I. Jordan. 2003. "Latent Dirichlet Allocation." Journal of Machine Learning Research 3 (Jan): 993-1022.

Blumelhuber, Benedikt, and Benedict Langer. 2026. "Generative AI Competencies in Demand: A Job Market Analysis of Knowledge, Skills, and Abilities."

Botero, David, and IAPP Westin Research Center. 2025. "US State Privacy Legislation Tracker." IAPP Resource Center. Last updated November 24, 2025.

Bushey, Kayla, and Saz Kanthasamy. 2024. "Ghost Jobs: The Phantom Hiring Trend with Startling Data Privacy Implications." IAPP. https://iapp.org/news/a/ghost-jobs-the-phantom-hiring-trend-with-startling-data-privacy-implications.

Carnegie Mellon University, Privacy Engineering and AI Governance Program. 2025. "Privacy Engineering Careers." https://privacy.cs.cmu.edu/masters/careers/. Accessed November 30, 2025.

Carnevale, Anthony P., Tamara Jayasundera, and Dmitri Repnikov. 2014. "Understanding Online Job Ads Data." Technical report. Georgetown University Center on Education and the Workforce. https://cew.georgetown.edu/wp-content/uploads/2014/11/OCLM.Tech_.Web_.pdf.

Cisco Systems, Inc. 2025. "Cisco 2025 Data Privacy Benchmark Study." https://www.cisco.com/c/dam/en_us/about/doing_business/trust-center/docs/cisco-privacy-benchmark-study-2025.pdf. Accessed October 27, 2025.

Dawson, Nikolas, Marian-Andrei Rizoiu, Benjamin Johnston, and Mary-Anne Williams. 2019. "Adaptively Selecting Occupations to Detect Skill Shortages from Online Job Ads." arXiv preprint. https://arxiv.org/abs/1911.02302.

Debortoli, Stefan, Oliver Muller, and Jan vom Brocke. 2014. "Comparing Business Intelligence and Big Data Skills: A Text Mining Study Using Job Advertisements." Business & Information Systems Engineering 6 (5): 289-300.

Demann, Yunique. 2025. "The New Triad of AI Governance: Privacy, Cybersecurity, and Legal." ISACA, @ISACA Newsletter. https://www.isaca.org/resources/news-and-trends/newsletters/atisaca/2025/volume-6/the-new-triad-of-ai-governance-privacy-cybersecurity-and-legal.

EDPS (European Data Protection Supervisor). 2014. "IPEN: Internet Privacy Engineering Network." https://www.edps.europa.eu/data-protection/ipen-internet-privacy-engineering-network_en.

Egger, Roman, and Jie Yu. 2022. "A Topic Modeling Comparison between LDA, NMF, Top2Vec, and BERTopic to Demystify Twitter Posts." Frontiers in Sociology 7: 886498.

ENISA (European Union Agency for Cybersecurity). 2022. "Data Protection Engineering." Technical report. January 2022. https://www.enisa.europa.eu/publications/data-protection-engineering.

European Commission. 2024. "AI Act: Shaping Europe's Digital Future." https://digital-strategy.ec.europa.eu/en/policies/regulatory-framework-ai.

European Parliament and Council of the European Union. 2016. "Regulation (EU) 2016/679 (General Data Protection Regulation)." Official Journal of the European Union L 119: 1-88. https://eur-lex.europa.eu/eli/reg/2016/679/oj/eng.

Fennessy, Caitlin. 2019. "Study: An Estimated 500k Organizations Have Registered DPOs across Europe." IAPP. https://iapp.org/news/a/study-an-estimated-500k-organizations-have-registered-dpos-across-europe.

Friedman, Raymond A., and Joel Podolny. 1992. "Differentiation of Boundary Spanning Roles: Labor Negotiations and Implications for Role Conflict." Administrative Science Quarterly 37 (1): 28-47.

Furnell, Steven. 2021. "The Cybersecurity Workforce and Skills." Computers & Security 100: 102080.

Graham, C. M. 2025. "AI Skills in Cybersecurity: Global Job Trends Analysis." Information & Computer Security 33 (5): 673-689.

Grootendorst, Maarten. 2022. "BERTopic: Neural Topic Modeling with a Class-Based TF-IDF Procedure." arXiv preprint arXiv:2203.05794.

Grybauskas, Andrius, Mantas Vilkas, Morteza Ghobakhloo, Mantas Lukauskas, and Viktorija Sarkauskaite. 2024. "Industry 4.0 Job Posting Profiles, Skills and Skills Impact on Salaries: A Natural Language Processing Approach." SSRN Working Paper 4755625. https://papers.ssrn.com/sol3/papers.cfm?abstract_id=4755625.

Gstrein, Oskar J., and Anne Beaulieu. 2022. "How to Protect Privacy in a Datafied Society? A Presentation of Multiple Legal and Conceptual Approaches." Philosophy & Technology 35 (1): 3.

Gurcan, Fatih, Ahmet Soylu, and Akif Quddus Khan. 2025. "Towards a Sustainable Workforce in Big Data Analytics: Skill Requirements Analysis from Online Job Postings Using Neural Topic Modeling." Sustainability 17 (20): 9293.

Gurses, Seda, and Jose M. Del Alamo. 2016. "Privacy Engineering: Shaping an Emerging Field of Research and Practice." IEEE Security & Privacy 14 (2): 40-46.

Gurses, Seda, Carmela Troncoso, and Claudia Diaz. 2011. "Engineering Privacy by Design." Paper presented at the Conference on Computers, Privacy and Data Protection (CPDP), Brussels, January 2011.

Heimes, Rita, and Sam Pfeifle. 2016. "Study: GDPR's Global Reach to Require at Least 75,000 DPOs Worldwide." IAPP The Privacy Advisor.

IAPP (International Association of Privacy Professionals). 2019. "IAPP Hits 50,000 Members, Marking a Milestone for the Organization and Growth of the Privacy Profession." May 2, 2019. Accessed November 2, 2025.

IAPP (International Association of Privacy Professionals). 2024a. "Privacy Governance Report 2024." Technical report. November 2024. https://iapp.org/resources/article/privacy-governance-report/.

IAPP (International Association of Privacy Professionals) and Ernst & Young. 2017. "IAPP-EY Annual Privacy Governance Report 2017." https://iapp.org/media/pdf/resource_center/IAPP-EY-Governance-Report-2017.pdf.

IAPP (International Association of Privacy Professionals) and TRU Staffing Partners. 2024c. "2023 Privacy Professionals Salary Survey." International Association of Privacy Professionals. https://iapp.org/resources/article/salary-survey-summary.

IAPP (International Association of Privacy Professionals). 2024b. "Key Terms for AI Governance." Updated July 2024.

Iwaya, Leonardo Horn, M. Ali Babar, and Awais Rashid. 2023. "Privacy Engineering in the Wild: Understanding the Practitioners' Mindset, Organisational Aspects, and Current Practices." IEEE Transactions on Software Engineering 49 (9): 4324-4348.

Jelodar, Hamed, Yongli Wang, Chi Yuan, Xia Feng, Xiahui Jiang, Yanchao Li, and Liang Zhao. 2019. "Latent Dirichlet Allocation (LDA) and Topic Modeling: Models, Applications, a Survey." Multimedia Tools and Applications 78 (11): 15169-15211.

Khan, Hammad Rauf, and Yunfei Du. 2018. "What Is a Data Librarian? A Content Analysis of Job Advertisements for Data Librarians in the United States Academic Libraries." IFLA WLIC, 1-9.

Kilhoffer, Zachary, Devyn Wilder, and Masooda Bashir. 2024. "Compliance as Baseline, or Striving for More? How Privacy Engineers Work and Use Privacy Standards." In 2024 IEEE European Symposium on Security and Privacy Workshops (EuroS&PW), 9-18. IEEE.

Kverne, Christopher Lukas, Federico Monteverdi, Agoritsa Polyzou, Christine Lisetti, and Janki Bhimani. 2025. "Course-Job Fit: Understanding the Contextual Relationship between Computing Courses and Employment Opportunities." In 2025 ASEE Annual Conference & Exposition.

Ludbey, Codee Roy, David J. Brooks, and Michael Coole. 2020. "Corporate Security Career Progression: A Comparative Study of Four Australian Organisations." Security Journal 33 (4): 531-551. https://doi.org/10.1057/s41284-019-00189-3.

Ma, Li, Ru Chen, Weigong Ge, Paul Rogers, Beverly Lyn-Cook, Huixiao Hong, Weida Tong, Ningning Wu, and Wen Zou. 2025. "AI-Powered Topic Modeling: Comparing LDA and BERTopic in Analyzing Opioid-Related Cardiovascular Risks in Women." Experimental Biology and Medicine 250: 10389. https://doi.org/10.3389/ebm.2025.10389.

McInnes, Leland, John Healy, and Steve Astels. 2017. "hdbscan: Hierarchical Density Based Clustering." Journal of Open Source Software 2 (11): 205.

McInnes, Leland, John Healy, Nathaniel Saul, and Lukas Grossberger. 2018. "UMAP: Uniform Manifold Approximation and Projection." Journal of Open Source Software 3 (29): 861. https://doi.org/10.21105/joss.00861.

NIST (National Institute of Standards and Technology). 2020. "NIST Privacy Framework: A Tool for Improving Privacy through Enterprise Risk Management (Version 1.0)." Technical report. https://nvlpubs.nist.gov/nistpubs/CSWP/NIST.CSWP.01162020.pdf.

NIST (National Institute of Standards and Technology). 2024. "NIST Privacy Workforce Taxonomy, Initial Public Draft." Technical report CSWP 38. November 2024. https://doi.org/10.6028/NIST.CSWP.38.ipd.

Ozyurt, Ozcan, and Ahmet Ayaz. 2024. "Identifying Cyber Security Competencies and Skills from Online Job Advertisements through Topic Modeling." Security Journal 37 (4): 1339-1359.

Pejic-Bach, Mirjana, Tanja Bertoncel, Maja Mesko, and Zeljko Krstic. 2020. "Text Mining of Industry 4.0 Job Advertisements." International Journal of Information Management 50: 416-431.

Persaud, Ajax. 2021. "Key Competencies for Big Data Analytics Professions: A Multimethod Study." Information Technology & People 34 (1): 178-203.

Radanliev, Petar. 2024. "Digital Security by Design." Security Journal 37 (4): 1640-1679. https://doi.org/10.1057/s41284-024-00435-3.

Rodriguez, Salvador. 2018. "Rise of the Data Protection Officer, the Hottest Tech Ticket in Town." Reuters, February 2018.

Rommetveit, Kjetil, and Niels van Dijk. 2022. "Privacy Engineering and the Techno-Regulatory Imaginary." Social Studies of Science 52 (6): 853-877.

Samarin, Nikita. 2024. "Measuring and Engineering Privacy Protections." PhD thesis, University of California, Berkeley.

Samarin, Nikita, Nandita Rao Narla, Liam Webster, and Daniel Smullen. 2025. "Defining Privacy Engineering as a Profession." Proceedings on Privacy Enhancing Technologies 2025 (4): 549-565.

Santos, Omar, and Peter Radanliev. 2024. Beyond the Algorithm: AI, Security, Privacy, and Ethics. Boston, MA: Addison-Wesley Professional.

Schwartz, Gabrielle, Joe Jones, and Uzma Chaudhry. 2024. "The Intersection of Privacy and AI Governance." May 2024.

Sentinella, Richard, Joe Jones, Ashley Casovan, Lynsey Burke, and Evi Fuelle. 2025. "AI Governance Profession Report 2025." International Association of Privacy Professionals and Credo AI. https://iapp.org/resources/article/ai-governance-profession-report/.

Szadeczky, Tamas, and Zsolt Bederna. 2025. "Risk, Regulation, and Governance: Evaluating Artificial Intelligence across Diverse Application Scenarios." Security Journal 38: 35. https://doi.org/10.1057/s41284-025-00495-z.

Todd, Peter A., James D. McKeen, and R. Brent Gallupe. 1995. "The Evolution of IS Job Skills: A Content Analysis of IS Job Advertisements from 1970 to 1990." MIS Quarterly 19 (1): 1-27.

TrustArc. 2023. "2023 Global Privacy Benchmarks Report." Technical report. TrustArc.

TrustArc. 2025. "2025 Global Privacy Benchmarks Report." Technical report. TrustArc. Based on survey responses from 1,775 professionals across industries and geographies.

Walczak, Basia, Hussein Abdulghani, Benjamin Kaplan, and Melanie Selvadurai. 2025. "From Compliance Cost to Competitive Edge: How Privacy Leaders Can Command the Executive Table." IAPP News.

Wang, Wenhui, Furu Wei, Li Dong, Hangbo Bao, Nan Yang, and Ming Zhou. 2020. "MiniLM: Deep Self-Attention Distillation for Task-Agnostic Compression of Pre-Trained Transformers." In Advances in Neural Information Processing Systems, vol. 33, 5776-5788. Curran Associates.

Watson, Cullen. 2023. "JobSpy." GitHub repository. https://github.com/speedyapply/JobSpy.

Wowczko, I. A. 2015. "Skills and Vacancy Analysis with Data Mining Techniques." Informatics 2 (4): 31-49. https://www.mdpi.com/2227-9709/2/4/31.

Yener, Ramazan, Muhammad Hassan, and Masooda Bashir. 2025. "Privacy Engineers on the Front Line: Bridging Technical and Managerial Skills." In Proceedings of the USENIX Conference on Privacy Engineering Practice and Respect (PEPR '25). Conference talk.

Zhang, Mike, Kristian Norgaard Jensen, Rob van der Goot, and Barbara Plank. 2022. "Skill Extraction from Job Postings Using Weak Supervision." arXiv preprint. https://arxiv.org/abs/2209.08071.